\documentclass[aps,prl,twocolumn,superscriptaddress,nofootinbib,
nobalancelastpage,nobibnotes]{revtex4}
\usepackage{epsfig}
\begin{document}

\title
{Effective mixing between clouds of different species of particles that are massless, or nearly massless.}

\author{R. F. Sawyer}
\affiliation{Department of Physics, University of California at
Santa Barbara, Santa Barbara, California 93106}

\begin{abstract}
 In collisions of intense beams, or clouds,
 coherent dynamics opens the way for fast transformations of light highly relativistic species into one another. Possible applications are suggested for situations, each  involving 
 two out of three of: photons, neutrinos, or gravitons. 
 \end{abstract}
\maketitle 
 \subsection{1. Introduction}

In the study of neutrinos in a supernova, there is a domain in which their densities are so high that neutrino-neutrino interactions can play the dominant role in determination of important correlations of $\nu$ flavors with $\nu$ energy spectra, this despite the fact that ordinary neutrino scattering is absolutely negligible in the systems in question \cite{raf1}-\cite{saviano}. This comes about through what we could describe as refractive effects: neutrinos taken to retain their individual momenta while energies change in a flavor-dependent way. 
The dynamics of the flavor evolution is then described by an effective Hamiltonian that operates totally in the space of flavor coordinates for each neutrino, as in addition do the familiar $\nu$-mass flavor terms. The Hamiltonian also depends parametrically on the angular distributions  $\nu$ cloud. 

The $\nu-\nu$ interaction from $Z_0$ exchange is effectively a point interaction. But when all of the neutrinos in the cloud overlap in space, then everybody interacts with everybody during the duration of the calculation.  At a mean-field level (which we define precisely below, and which serves in the supernova $\nu$ interactions) the theory leads to sets of coupled nonlinear equations for the density matrices in flavor space. The possible instabilities of these equations with respect to linearized perturbations play a decisive role in outcomes, the most dramatic of which is a sudden flavor exchange between clouds with differing energy 
\cite{rfs2}-\cite{mir}.

The time scale for these flavor transformations can be of the order $T\sim [n G_F]^{-1}$, where $n$ is the number density and $G_F$ is the Fermi constant.
But this last statement is subject to two {\it caveats}: 1) it holds only in parameter domains in which a flavor instability obtains (growing modes in a linearized 
system)\cite{rfs2}-\cite{mir}.  2)in addition the idealized initial state with occupation of pure flavor states can be at a point of unstable equilibrium. In the latter case a seed is available in the form of the flavor-mixing neutrino mass terms, which by themselves reflect a time scale of the order of hundreds of times longer than the 
$\nu-\nu$ coupling above, but which, in conjunction with the $\nu-\nu$, would lead to a transition time for supernova core $\nu$'s of order  
$T\sim [n G_F ]^{-1} \log (1 {\rm MeV}/m_\nu)$, if we started with a state with no flavor mixing.

Some of this lore will apply to other problems that involve dense clouds of highly relativistic particles. As an example we shall consider the possibility that an appreciable fraction of a very dense cloud of gravitons can transform into photons through a coherent mechanism similar to that driving the $\nu$ scenarios touched on above. And we suggest that somewhere in the future catalogue of extreme events 
the parameters that might enable this reaction to go forward.
It could seem wildly unlikely at first that the behavior of intensive classical gravitational radiation in some process could have a relation to neutrino calculations that involve a cloud of Fermions nearly thermalized in energies (but \underline{not} in the flavor space.)
And it could seem equally unlikely that, even if that were the case, the simplest treatment would be to start at the single graviton level.
But in the flavor space itself, where the calculations take place and flavor means ``being a graviton" or ``being a photon", the basic considerations are the same in the two realms.

Later we consider two different possible mechanisms for instigating a sudden transformation of a macroscopic fraction of the gravitons present into photons. The first is that the macroscopic motions of bulk matter that produce classical gravitational waves also produce some photons in the same modes, owing, e.g., to some magnetic fields locked into the matter. Then a classical purely gravitational calculation could presumably reveal the instability, and the classical photon ``seed" stimulate a large scale `` flavor" turnover. The second mechanism would stay completely in the gravitational domain, that is, take the photons to have only gravitational interactions, but to add a quantum effect. Normally we probably expect quantum effects in a system of $N$ coherent bosons to be a correction to classical results that are of order $N^{-1}$, and $N$ for the our purposes here is immense. But in situations dominated by the analogue to the ``flavor instability" we find instead a macroscopic transition time proportional to 
$\log[N]$. 

To investigate this possibility we begin at the two graviton level with the process ${\rm gr +gr} \rightarrow \gamma + \gamma$ where here and henceforth ``gr" stands for graviton. And for the moment we take the graviton initial state to be a plane wave with moment $\vec q$  and one of momentum $\vec k=-\vec q$. All of the relativity that goes into the present work is subsumed in the forward (or backward , depending on definitions) amplitude, $M$, for the helicity transition , $\{2,2\}\rightarrow \{1,1\}$  which is given in ref. \cite{grav2} in terms of Mandelstam variables, $s,t,u$ for the reaction ${\rm gr +\gamma \rightarrow gr + \gamma}$ as
$M=G s^2 t^{-1}$. For our crossed reaction we evaluate at $s=- t=2{\bf |q| \,|k|}(1-\cos \theta_{\bf q,k})$, $u=0$ obtaining
 \begin{eqnarray}
 M=-16 \pi G |{\bf  q| |k| (1-cos \theta_{q,k}})\,,
 \label{grint}
 \end{eqnarray}
where G is the gravitational constant. The result (\ref{grint}) applies equally to the other reaction channel  $ \{- 2,-2 \} \rightarrow \{-1,-1 \}$ that has non-vanishing amplitude. From (\ref{grint}) we shall build an effective Hamiltonian in terms of the creation and annihilation operators for the respective particles in the interaction. 
In the end we shall find that the wave-lengths and intensities of gravitational waves in the vicinity of the presumed black hole merger in the first LIGO event lead to an estimate graviton number density that is somewhat short of the density needed for the conversion process, given the time interval available (we consider a factor $10^{-2}$ in number density  ``somewhat short''). But considering the history whereby ever more extreme conditions
come to the fore it appears to us that that a venue with requisite parameters will come along someday. And the basic dynamics can apply to systems with other particles of very small mass.  Suggestions for the latter abound in the literature, most recently in connection with some dark matter speculations.

In the sections that follow we discuss a number of generic aspects of the interactions of dense clouds of massless particles, first explaining some of the above assertions in the simplest cases, then following with general approaches that support the results and also  give clues as to how things work when the prototype of two mono-directional beams is replaced by continuous angular distributions.

 \subsection{ 2. Simple guess for a wave-function }

We take $a_{\bf _q}, b_{\bf_ k}, c_{\bf_ q}, d{\bf_k}$ as the respective annihilators of momentum states for the different species. In this section we use a center of mass system, where ${\bf k=- q}$ but we keep the notations separate for later use when we turn to angular distributions. 
We take the reaction $A+B \rightarrow C+D$ to stand for,
\begin{eqnarray}
{\rm [gr]}_{\bf q} +{\rm [gr]}_{\bf k} \rightarrow \gamma_{\bf q} +\gamma_{\bf k} \,.
 \end{eqnarray}

At the origin in time we begin with our two clashing graviton waves, overlapping in space, in a coherent state with wave function 
\begin{eqnarray}
|\Psi (t=0) \rangle =Z e^{\sqrt{ N_A} [a_{\bf q}^\dagger-a_{\bf q}]} 
e^{\sqrt {N_B} [b_{\bf k}^\dagger-b_{\bf k}]} |0\rangle \,.
\label{coh}
\end{eqnarray}
where $Z$ is normalization.
$N_A$ and $N_B$ are the average number of particles in the respective beams. We calculate time evolution in a box of volume $V$ with periodic boundary conditions, with the times not to exceed the length of the box. In accord with our introductory remarks, we work in a time domain that is very short compared to a scattering time as computed from the cross-section;
the interactions that matter over this time period are just those in which the momenta stay the same but the labels that signify species change. For
$A({\bf q})+B({\bf k})\rightarrow C({\bf q})+D({\bf k})$, the relevant part of an effective interaction is taken as
\begin{eqnarray}
H_{\rm eff}=-i V^{-1} g_{\bf q,k} \Bigr [ \,c^\dagger_{\bf p}~ a_{\bf p} ~d_{\bf q}^\dagger ~b_{\bf q} - \,c_{\bf p}~ a^\dagger_{\bf p} ~d_{\bf q} ~b^\dagger_{\bf q}\Bigr ]\,,
\label{hamclass}
\end{eqnarray}
where in the actual gravitational case we shall take $g_{q,k}=G  (1-\cos \theta_{q,k})$, from (\ref{grint}), recognizing that a factor $\bf |q|\,| k| $ is to be removed from the amplitude in order to get the effective Hamiltonian.
We introduce time dependent linear combinations,
\begin{eqnarray}
\tilde a(t)=a_{\bf p} \cos \theta_A(t)- c_{\bf p} \sin\theta_A(t)\,,
\nonumber\\
\tilde c(t)=a_{\bf p} \sin \theta_A(t)+c_{\bf p} \cos \theta_A(t)\,,
\nonumber\\
\tilde b(t)=b_{\bf q}\cos \theta_B(t)-d_{\bf q}\sin\theta_B(t)\,,
\nonumber\\
\tilde d(t)=b_{\bf q} \sin \theta_B(t)+d_{\bf q}\cos \theta_B(t)\,,
\label{trans}
\end{eqnarray}
with $\theta_A(0)=0$, $\theta_B(0)=0$, and the time dependence of the $\theta$'s  to be determined.

Then we take an {\it ansatz} for the time dependent wave-function of the complete system,
\begin{eqnarray}
|\Psi(t) \rangle =\mathcal{C} e^{\sqrt N_A [\tilde a(t)^\dagger-\tilde a(t)]}  e^{\sqrt N_B [\tilde b(t)^\dagger-\tilde b(t) ]}|0\rangle \,.
\label{anz}
\end{eqnarray} 
This describes a coherent state such that as $\theta (t)$ changes, the A beam gradually acquires an ``C" part, and the B beam a ``D" part, representing a mixing with photons.
 
Now we demonstrate that there is a choice of time dependence for $\theta(t)$ such that  $|\Psi(t) \rangle  $ is a solution of the Schrodinger equation based on (\ref{hamclass}) when we include only leading powers of $N_A^{-1}, N_B^{-1}$. The steps in a completely heuristic approach are:
1) to write the Hamiltonian (\ref{hamclass}) in terms of the $\tilde a(t),\tilde b(t), \tilde c(t),  \tilde d(t)$ operators;
2) then to recognize that at time $t$ the leading contributions from $\tilde a,\tilde b$ are given by the replacements
\begin{eqnarray}
\tilde a=\sqrt{N_A}~, ~\tilde b=\sqrt{N_B}\,,
\label{cno}
\end{eqnarray}
both c-numbers, leaving the instantaneous time development in the hands of $\tilde c$ and $\tilde d$.
We substitute in  the Hamiltonian (\ref{hamclass}) 
operators, and retaining only the leading term for large $N$ in the result gives, 
\begin{eqnarray}
&H_{\rm eff}=-ig V^{-1}(\sqrt {N_A N_B} \Bigr[\sqrt{N_A}\sin(2\theta_B) (\tilde c^\dagger-\tilde c) 
\nonumber\\
&+\sqrt{N_B}\sin(2\theta_A) (\tilde d^\dagger -\tilde d) \Bigr ])\,.
\label{ham2a}
\end{eqnarray}
\subsection{3. Solution}
To calculate the evolution from just the leading part of the interaction, $H_{\rm eff}$ , (\ref{ham2a}) we note that

\begin{eqnarray}
(\tilde c^\dagger-\tilde c) |\Psi(t) \rangle =-N_A^{-1/2}{d \over d \theta_A} |\Psi(t) \rangle \,,
\end{eqnarray}
\begin{eqnarray}
 (\tilde d^\dagger-\tilde d) |\Psi(t) \rangle =-N_B^{-1/2} {d \over d \theta_B} |\Psi(t) \rangle\,,
\end{eqnarray}
and we can now express $ H_{\rm eff} |\Psi \rangle$ as
\begin{eqnarray}
&H_{\rm eff}|\Psi(t) \rangle=ig V^{-1}\Bigr[N_B \sin(2\theta_B) {\partial \over \partial \theta_A} |\Psi(t) \rangle
\nonumber\\
&+ N_A \sin(2\theta_A){\partial \over \partial \theta_B} |\Psi(t) \rangle\,.
\label{comp}
\end{eqnarray}
But now, since we are taking all time dependence to be embodied in $\theta_A$ and $\theta_B$, we have,
\begin{eqnarray}
{\partial \over \partial t}|\Psi(t) \rangle={\partial \theta_ A \over \partial t} {\partial \over \partial  \theta_A}|\Psi (t) \rangle+
{\partial \theta_ B \over \partial t} {\partial \over \partial  \theta_B}|\Psi (t) \rangle\,.
\end{eqnarray}
 Comparing (\ref{comp}) with the Schrodinger equation based on the Hamiltonian (\ref{ham2a}),   and equating the coefficients of
$[\partial / \partial \theta_B]|\Psi (t) \rangle$, $ [\partial / \partial  \theta_A]|\Psi (t) \rangle$, respectively, we see that the ansatz (\ref{anz})
   for the wave function obeys the Schrodinger equation if,
  \begin {eqnarray}
{d \theta_A \over dt}=g V^{-1}  N_B \sin(2 \theta_B)\,,
\label{eom1}
 \end{eqnarray}  
 and
  \begin {eqnarray}
{d \theta_B \over dt}=g V^{-1}N_A  \sin(2 \theta_A) \,.
\label{eom2}
\end{eqnarray}

If we begin at $t=0$ with initial values,  $\theta_A(0)=\theta_B(0)=0$, the system goes nowhere at all. If instead we choose very small values of order $\epsilon$ for either or both of the initial angles, then the configuration would transform into a 50-50 mixture of photons and gravitons in a time that is proportional to $\log \epsilon^{-1}$. In this latter case we would be beginning in a state with a small number of photons in the respective modes $N_{C}=(\epsilon_A)^2 N_A $, $N_{D}=(\epsilon_B)^2 N_B$. 
There may be some physics in seeds of this kind. 
Consider a process that produces a large flow of coherent gravitons by moving large masses around. If the matter had, for example, a small locked-in magnetic field, it would produce a small number of seed photons with exactly the wave number as a gravitational wave produced in the same process. 

For the case $N_A=N_B=N$ we define 
 \begin{eqnarray}
 \zeta(t)=\cos (2\theta)=N^{-1}\langle (a^\dagger a-c^\dagger c)\rangle \,,
 \label{zeta}
  \end{eqnarray}
which in our application is the difference between the probability that a graviton has remained a graviton and the probability that it has become a photon, and note that $\zeta (0)=\epsilon^2$, from (\ref{anz}).
In fig.1 we plot  solutions for $\zeta(t)$ for a range of values of $\epsilon$, with time in units of $(N g)^{-1} V = (n g)^{-1}$, where $n$ is the number density in the cloud. 
 
 \begin{figure}[h] 
 \centering
 \includegraphics[width=2.5 in]{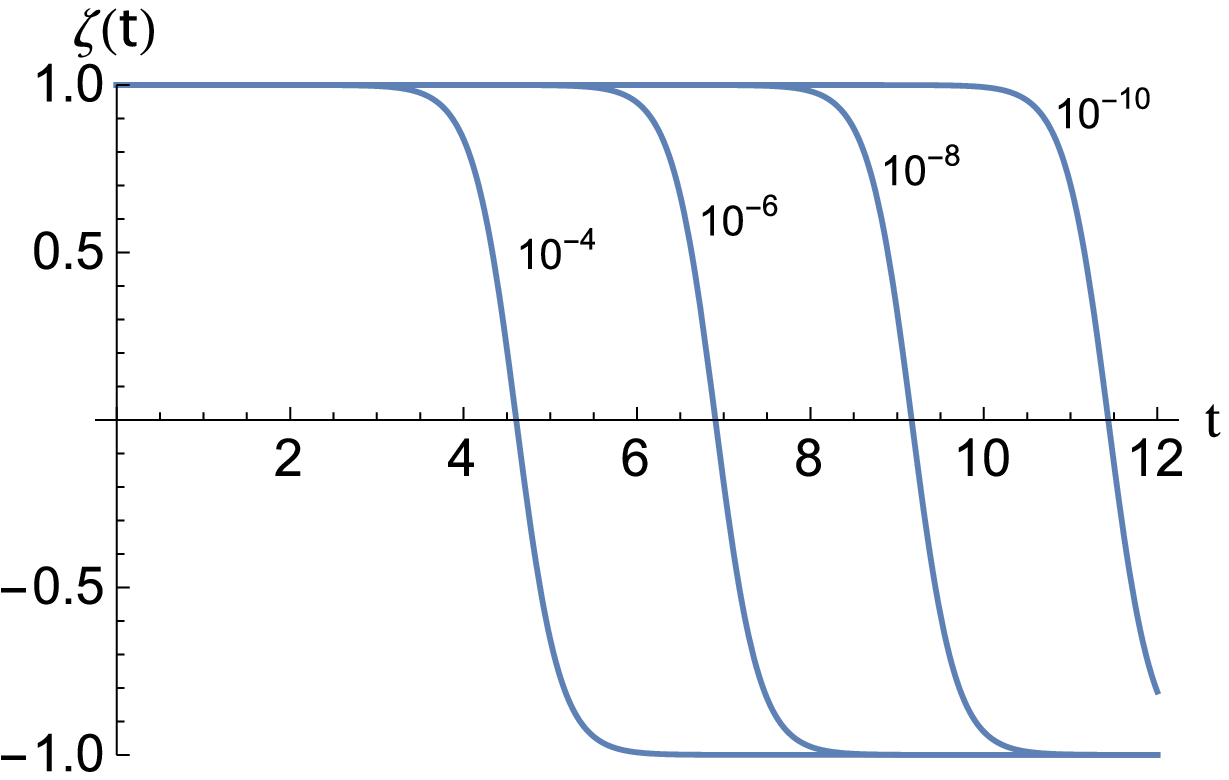}
 \caption{ \small }
Flavor evolution for the seeded model showing the results for different values of the initial seed $\zeta(0)$ ranging from $10^{-4}$ to $10^{-10}$. The long time result
for each case is that all initial gravitons have become photons. Time is in units $(n  g)^{-1}$.
\label{fig. 1}
 \end{figure}

The equal spacings in $\zeta$ as the size of the seeds $\theta (0)$ are decreased geometrically indicates a flavor turn-over time that is proportional to $\log[\zeta(0)^{-1}]$. This appears to be a purely classical result; since the only way $\hbar $ would enter the result is through particle numbers $N$, which when re-expressed in field strength are proportional to $ \hbar ^{-1}$, but $\epsilon_C=(N_C/N_A)^{1/2}$ is the ratio of such numbers.
We might speculate that taking $N_C=N_D=1$ gives a quantum estimate for the rate in the totally unseeded case. In the next section we more or less confirm that this is correct. 

\subsection{ 4. Numerical solution for a simplified, unseeded, and completely quantum case}

In the preceding we defined operators $a_{\bf q}$, $b_{\bf k}$ that annihilate single gravitons in the respective momentum states, and $c_{\bf q}$, $d_{\bf k}$ that do the same for photons, and we considered just the two spatial modes ${\bf k}$ and ${\bf g}$. Now we extend to $4N$ modes, for the moment not specifying their spatial momenta. The basic operators are taken as the set of annihilators
$\{a_i, b_i,c_i,d_i\}$ where the index $i$ runs from 1 to N.  We define the bilinears,
\begin{eqnarray}
\sigma_{\bf q}^{(+)}=a^\dagger_{ i} \,c_{ i}~,~ \sigma_{i}^{(-)} =[\sigma^{(+)}_{ i} ]^\dagger~ ,~\sigma^{(3)}_{i}=a^\dagger_{ i} a_{i}- 
c^\dagger_{i} c_{i}\,,
\nonumber\\
\tau^{(+)}_{j}=d^\dagger_{j} \,b_{j}~,~ \tau^{(-)}_{j} =[\tau^{(+)} _{j} ]^\dagger~ ,~\tau^{(3)}_{j} =d^\dagger_{j} d_{\bf k}- b^\dagger _{j}b_{j}\,,
\label{bil}
\end{eqnarray}
which have Pauli matrix commutation rules (defining $\sigma_+=[\sigma_1+ i \sigma_2)/2]$ etc.).
The Hamiltonian is the multimode generalization of (\ref{hamclass})
\begin{eqnarray}
H=i V^{-1} \sum _{i, j} g_{i, j} [\sigma^{(+)}_i \tau^{(-)} _j- \sigma^{(-)}_i \tau^{(+)} _j]\,.
\label{ham3}
\end{eqnarray}

In (\ref{ham3}) we first take $ g_{\bf q, k} =g$, a single coupling strength between each pair of states. We have good evidence that small fluctuations of the strengths change results hardly at all. Thus the approximation should serve for beams that are reasonably monochromatic and collimated.  In the initial state we take occupancies of unity for species annihilated by $a_i$ and $b_i$, and zero for those of annihilated by $c_i$ and $d_i$. We introduce collective variables $ \vec \sigma=\sum_j^{N_A}\vec \sigma_j$,
$ \vec \tau=\sum_j^{N_B}\vec \tau_j$ and write,

\begin{eqnarray}
H_S=g V^{-1} [\sigma^{(+)} \tau^{(-)}+\sigma^{(-)} \tau^{(+)}] \,,
\label{hs}
\end{eqnarray}
where the commutation rules are $[\sigma^{(+)},\sigma^{-}]=\sigma^{(3)}$,  $[\sigma^{(+)},\sigma^{(3)}]=2 \sigma^{(+)}$, etc. The initial state $\Psi[t=0]$ obeys $ \sigma^{(3)}\Psi=N_A \Psi$ ,  $ \tau^{(3)}\Psi=-N_B \Psi$. 
 Now naming the flavors of individual particles as ``spins" we find that
 $\vec \sigma \cdot \vec \sigma /4=(N_A/2+1)N_A/2$, which is conserved, and similarly  
for the B state, upon replacing $\vec \sigma$ by $\vec \tau$ and $N_A$ by $N_B$. The quantity $\sigma^{(3)}+\tau^{(3)}$ is also conserved.
 The calculation is particularly simple when $N_A=N_B=N$. Then we have just $N+1$ states contributing to the evolution
 and the relevant matrix elements of the operators in $H$ are,
\begin{eqnarray}
\langle i-1 |\sigma^{(-)} \tau ^{(+)} |i\rangle=i (N-i+1)\,,
\nonumber\\
\langle i |\sigma^{(+)} \tau ^{(-)} |i-1\rangle=i (N-i+1)\,.
\label{equal}
\end{eqnarray}

We solve for this wave function $\Psi(t)$ in the $N+1$ dimensional subspace and again plot $\zeta(t)$
Fig.2 plots the evolution of $\zeta(t)=N^{-1}\langle \sigma_3(t)\rangle$
for values $N$ ranging from 16 to 1024, equally spaced by factors of four.  
 \begin{figure}[h] 
 \centering
\includegraphics[width=2.5 in]{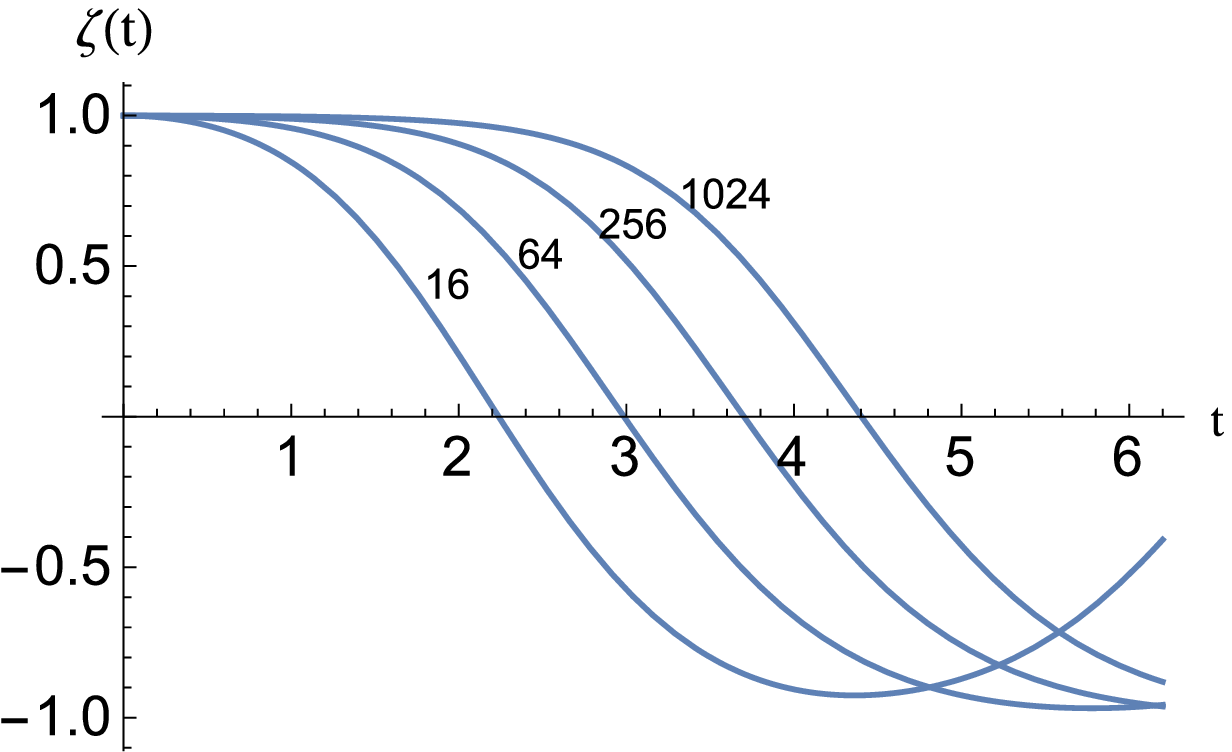}
 \caption{ \small }
Flavor evolution for one of the beams for the $N_A=N_B=N$ case, with $N$ ranging from 16 to 1024. Time is in units $(n g)^{-1}$. $\zeta=1$
indicates 100\% $\gamma$'s; 
\label{fig. 2}
\end{figure}

The times for the $\zeta=0$ intercepts are given by $T=.65 (gn)^{-1} \log N$, showing the expected logarithmic dependence. 
 The turnover times of order $(gn)^{-1} \log N$ displayed in fig. 2 would confirm the conjecture at the end of the last section that the quantum result 
for the unseeded case would be of the order of the result of the seeded result of section 2 if the seeds were chosen to comprise a single photon.

\subsection{5. Mean-field theory}

There are many questions that one can raise about the above development. For example, whether retaining only the $N^{-1/2}$ terms in 
$H_{\rm eff}/N^2$ as in (\ref{ham2a}) is legitimate.  (We omit the irrelevant leading term of order unity as $N\rightarrow \infty$.) Of course we shall take $N$ to be enormous.  But we are looking not just for the short-term change in the wave-function, which is indeed of order $(H_{\rm eff}t) ^2$, but for a long time development in which the wave-function drastically changes its nature. 

A more focused question is that of what happens when we depart from the two-ray (or clashing beams) model to have angular distributions instead of plane waves. At best we can expect some diminution of effect by virtue of the $(1-\cos\theta_{\bf q,k})$ factor in the input scattering amplitudes. 

We can approach  these questions in a mean-field (MFT) limit. At the same time we connect to previous work on neutrinos, where the equivalent to our mean-field theory is operative \cite{raf3}. We begin from the equations of motion under the action of the interaction (\ref{ham3}) for the bilinears defined in (\ref{bil}),

\begin{eqnarray}
&  V\dot \sigma^{(+)}_{\bf q}= \sigma^{(3)}_{\bf q}\sum_{\bf_k} g_{\bf q,k} \, \tau^{(+)}_{\bf k}~,~
  V\dot \tau^{(+)}_{\bf q}=- \tau^{(3)}_{\bf q}\sum_{\bf_k} g_{\bf q,k} \,\sigma^{(+)}_{\bf k}  ~, 
  \nonumber\\
  &V \dot \sigma^{(3)}_{\bf q}=\sum_{\bf_k} g_{\bf q,k}  (\sigma^{(-)}_{\bf q}\, \tau^{(+)}_{\bf k}-\sigma^{(+)}_{\bf q}\, \tau^{(-)}_{\bf k} )~,
  \nonumber\\
   &V \dot \tau^{(3)}_{\bf q}=-\dot \sigma^{(3)}_{\bf q}\,.
    \label{eom}
\end{eqnarray}

Mean field theory, in the present problem, is the assumption that in the equations of motion (\ref{eom}) we can replace each variable $\vec \sigma, \vec \tau$ by its expectation value in the medium. Thus we demand, e.g., that $\langle \sigma^{(3)}_{\bf q} \tau^{(+^)}_{\bf k}\rangle =\langle \sigma^{(3)}_{\bf q } \rangle\langle \tau^{(+)}_{\bf k}\rangle$. In what follows, we shall always choose the initial time wave function to satisfy these conditions, but at later times it is only an approximation that holds (if at all) in the $N\rightarrow \infty$ limit.

For the single mode case that we solved in the last section
we have simply,
\begin{eqnarray}
i  \dot \sigma^{(+)}=g V^{-1}\sigma^{(3)} \tau^{(+)} \,,
\nonumber\\
i  \dot \tau^{(+)}=g V^{-1} \tau^{(3)} \sigma^{(+)}  \,,
\nonumber\\
i  \dot \sigma^{(3)}=g V^{-1}[\tau^{(+)}  \sigma^{(-)}-\tau^{(-)}  \sigma^{(+)}]\,,
\label{mf1}
\end{eqnarray}
where the variables $\sigma$ and $\tau$ are now just numerical functions of $t$ rather than operators in our original space and the initial conditions are $\sigma_3=N$, $\tau_3=N$. If we take the initial mixings, $\sigma^{(+)}(0)=0$, $\tau^{(+)}(0)=0$, then nothing ever happens. 
Taking  $\sigma^{(+)}(0)=\epsilon$, $\tau^{(+)}(0)=\epsilon$, we can compare the resulting MFT time development with one of the plots shown for the full theory in fig. 2.;  we choose an initial $\epsilon$ 
that fits the behavior at the earliest times; which turns out to be $\epsilon\approx 1$.
The results are shown in fig. 3 for the case $N=512$.
 \begin{figure}[h] 
 \centering
\includegraphics[width=2.5 in]{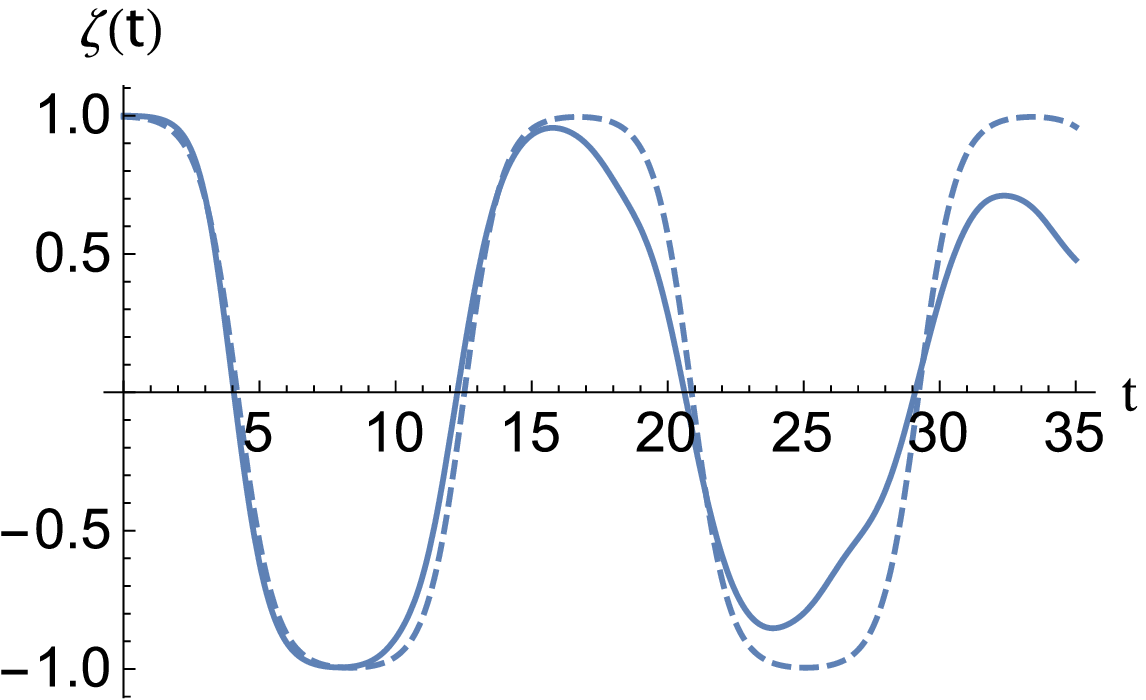}
 \caption{ \small }
Plot of the solution for the value $N=512$, as plotted in fig.2 but here extended in time, with the dashed curve showing the mean-field calculation.
\label{fig. 3}
\end{figure}
We see that the periodic MFT solution tracks the real solution well through the first half period, after which the oscillations in the real solution begin
to damp, while those in MFT remain maximal. These qualitative features appear to be the same for all $N$ up to $1024$, and, we hope, for $N=10^{20}$. In any case, we will provisionally
accept mean-field calculations as a way of extending the results of our one complete {\it ab initio} calculation, that of sec. 3, into domains that would otherwise be inaccessible to us in the full theory by virtue of their complexity.

The space momenta, which remain fixed, just supply a substrate for the dynamics in the flavor sector and the difference between Fermi and Bose systems is really inconsequential except in one respect: in the photon-graviton case with some tiny initial complement of photons we can have
$\langle c_{\bf q}\rangle \ne 0$, $\langle a_{\bf q}^\dagger \rangle \ne 0$, so that the bilinear,
$\langle \sigma^{(+)}_{\bf q} \rangle \ne 0$ at time zero can serve as the seed for the process. (In the conventional neutrino case the seed is provided instead by the ordinary oscillation term, as discussed in sec.1). This feature comes from the assumption of a coherent state, as could be produced
by macroscopic currents tied to motions of large masses. If our initial photons were thermal then there would be no essential Bose-Fermi difference.

When instead of a distribution of two clashing beams, the distribution is isotropic, we use (\ref{eom}) with a sample of ${\bf k}$'s isotropically distributed 
in angle and $g_{\bf q,k}=g (1-\cos \theta_{\bf q,k})$ from (\ref{grint}). Now the equations can easily be solved numerically. 
In fig. 4 we show the resulting evolution. Shown are the the results for the same choice of $N$ and initial values, but replacing  $(1-\cos \theta_{\bf q,k})$ by its average over solid angle. 
\begin{figure}[h] 
 \centering
 \includegraphics[width=2.5 in]{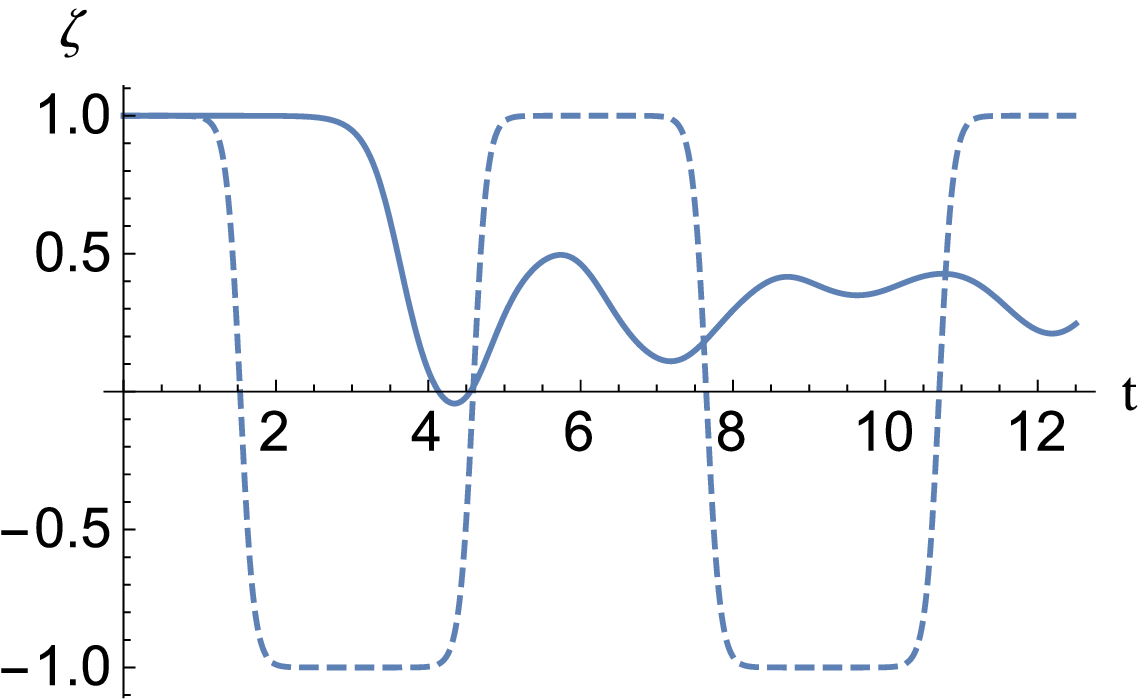}
\caption{ \small }
Evolution in mean-field approximation with just two clashing beams (dashed curve) of compared to a case in which the same number of particles
are distributed isotropically in two clouds with one of the initial clouds being pure flavor A and $\zeta(t)$ measuring the production of flavor
C, following definitions in text.
\label{fig. 4}
\end{figure}
We here see that spreading the same number of gravitons as in the beams case into an isotropic distribution lengthens the break time by about a factor of two, and leads to a gentler dive, and to a final state that when the calculation is extended to a greater time gives one half gravitational wave and one half
electromagnetic wave. 
\subsection{6. Instabilities}
Instabilities, which we have mentioned only in passing up at this point are at the heart of all of the results in this paper.
To explain the essence of this remark we go back to the very simplest mean-field model (\ref{hs}) adding one final term to the Hamiltonian,

\begin{eqnarray}
H_S=g V^{-1} [\sigma^{(+)} \tau^{(-)}+\sigma^{(-)} \tau^{(+)}+{1\over 2} \lambda \sigma^{(3)} \tau^{(3)}] \,,
\label{hs1} 
\end{eqnarray}
With the ensuing equations
\begin{eqnarray}
i  \dot \sigma^{(+)}=g V^{-1}[\sigma^{(3)} \tau^{(+)}  +\lambda \tau^{(3)} \sigma^{(+)}]\,,
\nonumber\\
i  \dot \tau^{(+)}=g V^{-1}[\tau^{(3)} \sigma^{(+)}  +\lambda \sigma^{(3)} \tau^{(+)}]\,,
\nonumber\\
i  \dot \sigma^{(3)}=g V^{-1}[\tau^{(+)}  \sigma^{(-)}-\tau^{(-)}  \sigma^{(+)}]\,.
\label{mf1}
\end{eqnarray}

When we take initial conditions, $\sigma^{(3)}=N$,  $\tau^{(3)}=-N$, $\sigma^{(+)}=\tau^{(+)}=0$, nothing at all happens. If we take tiny initial values
for $\sigma^{(+)}\,,\,\tau^{(+)}$ in linearized equations, there are exponential growth of these perturbations only if there are imaginary parts to 
the eigenvalues of the matrix
\newline

$ W=\left(
\begin{array}{cc}
 \lambda & 1 \\
 -1 &-\lambda \\
\end{array}
\right)$
\newline

When $0<|\lambda| <1$, then these eigenvalues are imaginary and when $|\lambda| \geq1$ they are real. In the latter case a tiny initial seed of , e.g., 
$\delta \sigma^{(+)}$ leads to nothing more than eternal small oscillations with the seed amplitude, while in the former we obtain the large scale flavor turn-overs of the models discussed above. 
This quenching effect of the $\lambda$ term is not mysterious by itself, since it measures the difference of the energies of purely gravitational interaction of a graviton with the medium and that of a photon with the medium. If this were too large it would necessarily overcome the 
transition matrix elements from $\sigma^{(-)}\tau^{(+)} $.
It appears that for our calculation, as long as the interactions are purely gravitational, we need not worry about there being a $\sigma^{(3)}\tau^{(3)}$ term, since the medium-dependent shift in energy should be exactly the same for the graviton as for the photon, as long as the only force is gravity.\footnote{This is in contrast to the neutrino-neutrino case. Here, e.g., in the two beam case with different initial flavors, the $\nu-\nu$ interaction coming from Z meson exchange, gives exactly $\lambda=1$, so there is no instability. Here the instabilities in recent literature result from more complex, and flavor-dependent, angular distributions \cite{va} -\cite{va3}.}

However, in our graviton-photon system once we have produced some critical density of photons the photon-photon interaction (coming from the virtual electron loop) is sure to produce energy shifts that would shut down the production mechanism.

If we return to the pure quantum calculation of sec. 4, but now with the added $\lambda$ term in $H_{\rm eff}$, then as expected from the mean-field behavior we obtain flavor turnover when $\lambda<1$ in a time $\approx (1-\lambda)^{-1}n^{-1} \log(N)$ time and none when $\lambda \ge 1$
 (The case $\lambda=1$ was solved analytically in ref. \cite{FL}.)  The difference between the two types of behaviors reflects the difference between stable equilibria and unstable equilibria in the related mean-field theories. 

There is a literature on condensed matter questions that shares an aspect of some of the above.  Given a mean-field description of a certain system with N components in unstable classical equilibrium, one asks what happens when quantum corrections drive it out of equilibrium.  Repeatedly it is found  \cite{va}-\cite{va3} that a ``quantum break time", at which the system begins to change suddenly, is proportional to  $\log N$.

\subsection{Numbers}
In the LIGO observation of the black hole merger reported in \cite{ligo}, the maximum luminosity in gravitational waves was reported to be
$\approx 3.6 \times10^{56}$ergs/ sec. at frequency of 250 Hz. Taking the implied number of gravitons in a sphere of one-wavelength, 
$\lambda_1$, radius yields a number density, $n \approx  10^{22}$MeV$^3$ in units with $\hbar=c=1$. Then defining $T$ as the time scale for traversing the diameter of the sphere and using $8 \pi G\approx 1.5 \times 10^{-43} {\rm (MeV)^{-1}}$, we estimate the quantity 
\begin{eqnarray}
\xi =8 \pi G n T \approx .01\,,
\label{ans1}
\end{eqnarray}
where $\xi \ge 1$ is required for flavor turn-over, and this is not even with in the increase of the turn-over time  by the logarithmic factor. Earlier
we found a simple factor of $\log N$ in the (quantum) case, or a factor of $\log N/ \log N_\gamma$ in the case in which there is an initial complement of photons. Thus we fall considerably short of large scale flavor turn-over. 

Of course the above times $T$ required to turn a sea of gravitons into any sizable percentage of photons with routine scattering mechanics is much shorter than times required for producing the same number using individual particle cross-sections times number densities. 
In the latter case the time necessary, under the above conditions, is longer by a factor
of order $G^{-1} \lambda_1^2 \approx 10^{85}$. This makes our estimate for the refractive process seem like a near-miss.

If nature provides systems with parameters that make our reactions actually happen we would need also to estimate the choking point in photon densities at which further  transformation is forbidden because of the photon-photon interactions in the medium; here we are moved out of the 
unstable region essentially by an analog to the $\lambda$ term in (\ref{mf1}). It turns out that the rough condition for this blocking to occur is
$8 \pi G n_{\rm gr} <.1 \alpha^2 E^2 m_e^{-4} n_{\gamma}$, where $E$ is the energy the mode, $m_e$ the electron mass, and $\alpha$ the fine-structure constant . For the parameters we assumed in the above example we find that $n_\gamma$ can even reach $n_{\rm gr}$.
 
 \subsection{8. Discussion}
 
 In this paper, we have discussed at some length a possible gravitational phenomenon that may not quite be realized in nature. In its purest form, where there is only gravitation coupling and there are no photons in the initial state, it is clearly a calculation involving quantum gravity, as shown by the $\log N$ factor in the answer, since $N$ as expressed in terms of field strength and wave-length brings in an $\hbar$. However when we replace this factor by $\log[N/N_\gamma]$ in the case with a photon seed it appears to be completely classical. 
 
This seems consistent, given a coupling term involving gravitational fields that is also quadratic in $F_{\mu\,\nu}$ and a photon seed from classical motions. To track our results classically would then involve taking one of these $F_{\mu\,\nu}$'s as the seed field and looking for the instability in the resulting linearized equation for $F_{\mu\,\nu}$, in the presence of the gravitational fields.
This appears to be a lot harder to execute than it is to begin with the photon-graviton scattering amplitudes and then to build the classical field answer the way we did. We could liken the situation to that of asking a quantum field theorist to calculate the Thompson scattering cross-section, which he or she, armed with the cookbook recipes, could do in five minutes in QFT. It would take this person a lot longer to do it {\it ab initio} from classical physics.

The methods of the present paper fall short of what would be needed even if the numerical estimates had been more propitious in the black-hole merger estimate of graviton number densities. We have omitted the background geometry in a region uncomfortably close to the horizon. All of our simulations were based on the system being in a box with periodic boundary conditions, in which the interaction is turned on everywhere all at once, and time is the only variable. 
In the neutrino prototype, where we envisioned a fairly steady flow for a second or so, it is then easy to replace the time dependent equation with a Liouville equation in space for steady flow \cite{raf3}.  The essential elements of the response to the instability are then the same as determined from the time-alone model, provided the densities and angular distributions of the cloud do not change much in the instability time. These conditions are not even nearly satisfied in the black-hole merger picture.  During the transition of species, if it happened at all, the densities would change by a lot. 

We also have not included angular distributions in our estimate, which was based on $\cos \theta=-1$. However in a mean-field simulation of an isotropic case (but with single occupancies for each mode) we found that the break time was only moderately longer than in the clashing beam case. 

Nevertheless, the end result of our estimate is eighty orders of magnitude greater than one would have estimated had one lacked knowledge of the instability in the refractive equations. Surely it is a problem worth looking into using more powerful methods.

There are other reasons for pursuing the central ideas of the paper:

1. Nature could well supply a venue in which the parameters required for ${\rm gr +gr \rightarrow \gamma+\gamma }$ are realized. 

2. Recent  theories in which  $m<10^{-20}$eV scalar particles provide the dark matter in the universe  \cite{peebles}-\cite{marsh} could be a playground for the methods of this paper. Here it is immaterial that the particles could be non-relativistic; the phases induced by these masses would stay inconsequential over vast time scales. The high number densities in these models make them promising venues.

3. Partially going back to the origins of these ideas in neutrino-world, we can ask about the reaction $\nu+\bar \nu \rightarrow \gamma +\gamma$. The arena would be the last moments of a neutron star's capture by a black hole, where we picture nearly all of the matter about to be accreted as comprising a relatively thin disk of matter with temperatures of a few MeV. The neutrino radiation from this disk provides a good fraction of the energy  loss in the capture. A bit above and along the axis of the disc neutrino beams can interact with each other. Our best estimates are that gammas could be created by this process only if anomalous magnetic moments are near their regions of exclusion by experiment, unless, of course, there were an anomalously large direct effective Hamiltonian for the $2\rightarrow 2$ process.

\end{document}